\newif\ifabstract
\abstracttrue
 \abstractfalse %Uncommenting this line gives the full version
\newif\iffull
\ifabstract \fullfalse \else \fulltrue \fi
\documentclass[11pt]{article}
\usepackage{rotating}
\usepackage{multirow}
\usepackage{amsfonts}
\usepackage{amssymb}
\usepackage{amstext}
\usepackage{amsmath}
\usepackage{xspace}
\usepackage{theorem}
\usepackage{graphicx}
\usepackage{url}
\usepackage{graphics}
\usepackage{colordvi}
\usepackage{colordvi}
\usepackage{subfigure}

\advance \oddsidemargin by -0.8in
\newcommand{\myparskip}{3pt}
\parskip \myparskip

\begin{document}

\title{Application of Econometric Data Analysis Methods to Physics Software\footnote{Presented at at 2016 IEEE NSS/MIC - Strasbourg, France, 29 October - 5 November, 2016}}

\author{Maria Grazia Pia\thanks{INFN at Genoa. Email: {\tt maria.grazia.pia@ge.infn.it}} \and Elisabetta Ronchieri\thanks{INFN CNAF at Bologna. Email: {\tt elisabetta.ronchieri@cnaf.infn.it}}}

\begin{titlepage}
\maketitle

\thispagestyle{empty}

\begin{abstract}
We report an investigation of data analysis methods derived from other disciplines, which we applied to physics software systems. They concern the analysis of inequality, trend analysis and the analysis of diversity. The analysis of inequality exploits statistical methods originating from econometrics; trend analysis is typical of economics and environmental sciences; the analysis of diversity is based on concepts derived from ecology and treats software as an ecosystem. To the best of our knowledge, this is an innovative exploration, as we could not find track of previous use of these methods in the experimental physics domains within the scope of the IEEE Nuclear Science Symposium. We applied these methods in the context of Geant4 physics validation and Geant4 maintainability assessment. 
\end{abstract}

\end{titlepage}

%---------------------------------------------------------------------------------------------------
%---------------------------------------------------------------------------------------------------
%---------------------------------------------------------------------------------------------------
%---------------------------------------------------------------------------------------------------
%----------------------------------------------------------------------------------------
\section{Introduction}\label{sec: intro}
%---------------------------------------------------------------------------------------------------
%---------------------------------------------------------------------------------------------------
%---------------------------------------------------------------------------------------------------
%---------------------------------------------------------------------------------------------------
%----------------------------------------------------------------------------------------

This presentation reports an innovative application of data analysis methods derived from 
disciplines other than physics, namely economy and ecology, to physics software problems.
They concern the analysis of inequality, trend analysis and the analysis of diversity. 
The analysis of inequality exploits statistical methods originating from econometrics; 
trend analysis is typical of economics and environmental sciences; 
the analysis of diversity is based on concepts derived from ecology and treats software as an ecosystem. 

The exploration of these methods is motivated by concrete requirements of ongoing projects
concerning Geant4 \cite{g4nim, g4tns} physics validation and Geant4 maintainability assessment; 
nevertheless, their scope of application in physics analysis is wider.

%---------------------------------------------------------------------------------------------------
%---------------------------------------------------------------------------------------------------
%---------------------------------------------------------------------------------------------------
%---------------------------------------------------------------------------------------------------
%----------------------------------------------------------------------------------------
\section{Inequality analysis}
\label{sec_econom}
%---------------------------------------------------------------------------------------------------
%---------------------------------------------------------------------------------------------------
%---------------------------------------------------------------------------------------------------
%---------------------------------------------------------------------------------------------------
%----------------------------------------------------------------------------------------

The need to detect and estimate inequality within a data sample may arise in a variety 
of physics scenarios.
We investigated its application in the context of evaluating software quality metrics \cite{ronchieri_chep2015},
where inequality analysis helps aggregating the sparse information associated with 
individual elements (files, classes) of a software package into a single variable, 
which summarizes the distribution of the measurements.

A set of statistical methods has been developed in the context of econometrics to 
identify and quantify inequality
\cite{ineq_handbook}. 
In their original context they are usually applied to evaluate the distribution of resources 
within a country.
For instance, the most common measure of inequality, the Gini index, 
is a measurement of the income distribution of a country's inhabitants. 
It is a number between 0 and 1, which is based on residents' net income;
it measures the gap between the rich and the poor, with 0 representing perfect
equality among the inhabitants and 1 representing perfect inequality.

In addition to the Gini index, we evaluated other inequality measures: they include 
the Ricci-Schutz coefficient (also known as Pietra index), Theil's entropy measure and Atkinson's measure.
We will discuss their specific characteristics and the role they play in the analysis of 
Geant4 maintainability in the full paper.

\begin{figure} 
\centerline{\includegraphics[angle=0,width=8cm]{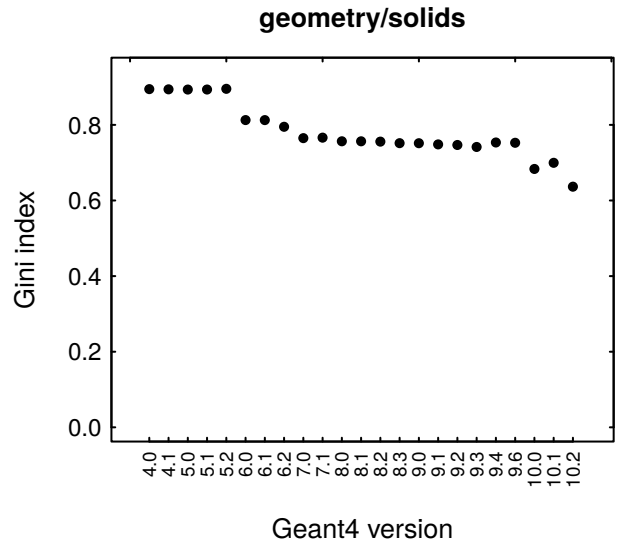}}
\caption{The Gini index calculated over Halstead Mental Effort measure in Geant4 
solids package, as a function of Geant4 version. This plot is the result of a preliminary analysis and is shown as an example of the outcome
of inequality analysis methods.}
\label{fig_gini}
\end{figure}

\begin{figure} 
\centerline{\includegraphics[angle=0,width=8cm]{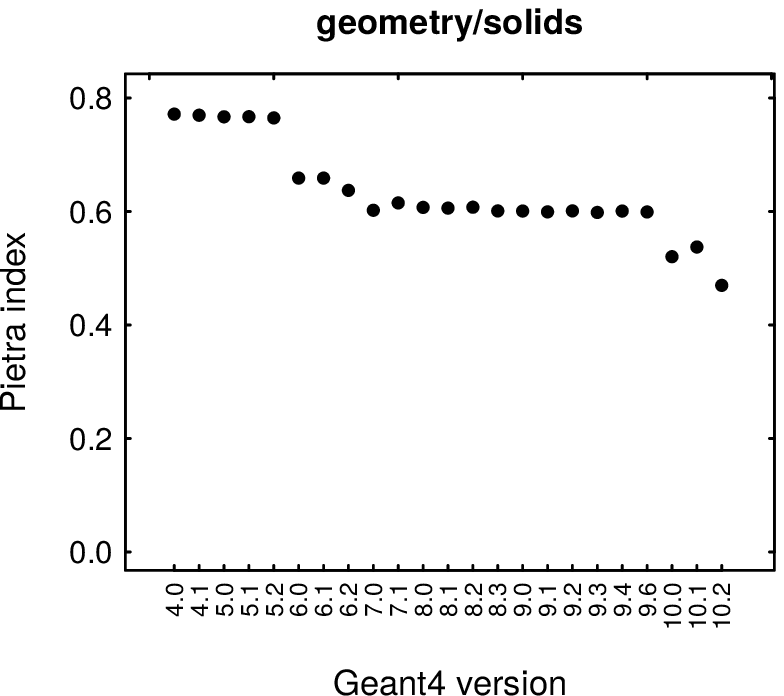}}
\caption{The Pietra index calculated over Halstead Mental Effort measure in Geant4 
solids package, as a function of Geant4 version. This plot is the result of a preliminary analysis and is shown as an example of the outcome
of inequality analysis methods.}
\label{fig_RS}
\end{figure}

\begin{figure} 
\centerline{\includegraphics[angle=0,width=8cm]{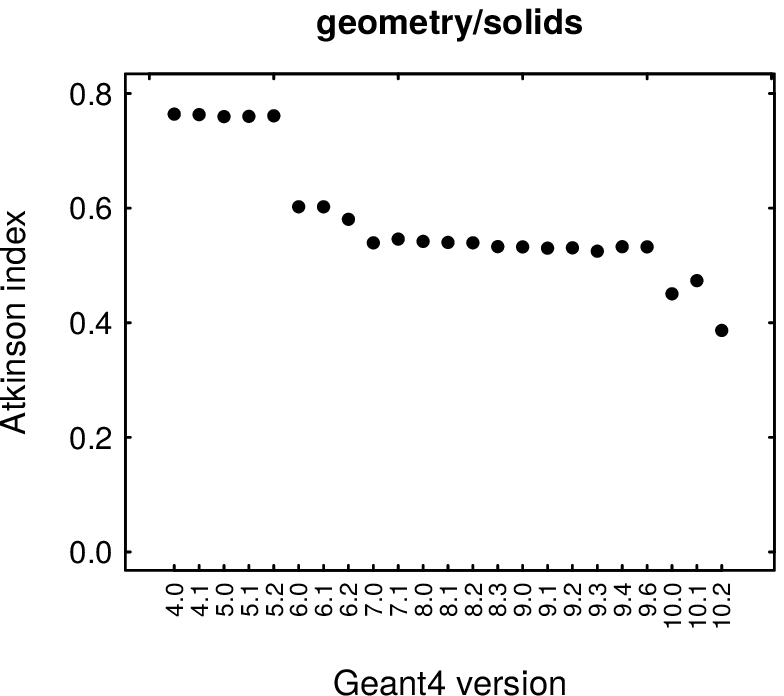}}
\caption{The Atkinson index calculated over Halstead Mental Effort measure in Geant4 
solids package, as a function of Geant4 version. This plot is the result of a preliminary analysis and is shown as an example of the outcome
of inequality analysis methods.}
\label{fig_atkinson}
\end{figure}

\begin{figure} 
\centerline{\includegraphics[angle=0,width=8cm]{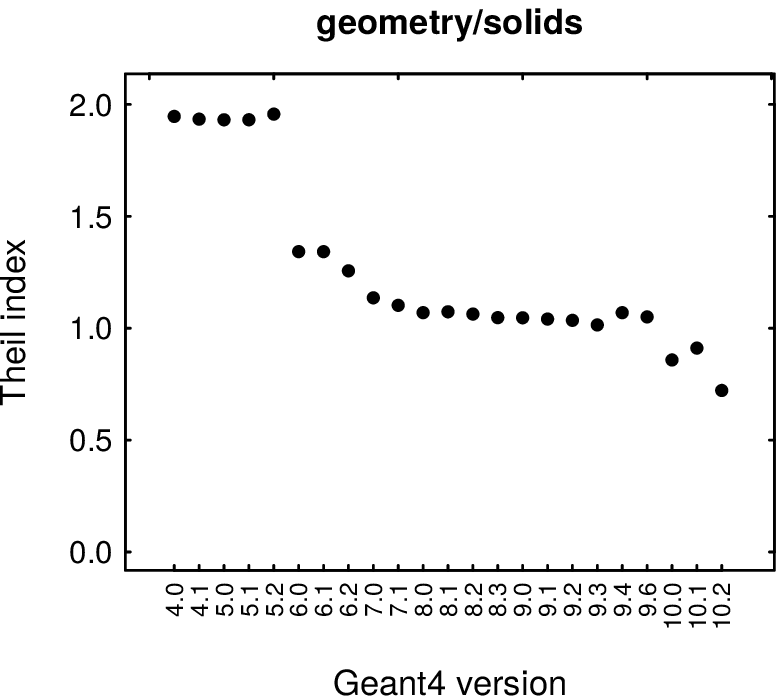}}
\caption{The Theil index calculated over Halstead Mental Effort measure in Geant4 
solids package, as a function of Geant4 version. This plot is the result of a preliminary analysis and is shown as an example of the outcome
of inequality analysis methods.}
\label{fig_theil}
\end{figure}

An example of inequality analysis is shown in Figs. \ref{fig_gini}-\ref{fig_theil}, which concern the Halstead  Mental Effort metric calculated over 
the \textit{solids} package of Geant4.
One can observe similarities across the various inequality measures, although their numerical values
are different.
%The large values of the Gini coefficient in both packages indicate large inequalities in the complexity of the software 
%across the classes belonging to these packages.

Currently, we are evaluating the application of inequality analysis to overcome some
limitations of inductivism in the context of simulation validation.

%---------------------------------------------------------------------------------------------------
%---------------------------------------------------------------------------------------------------
%---------------------------------------------------------------------------------------------------
%---------------------------------------------------------------------------------------------------
%----------------------------------------------------------------------------------------
\section{Trend analysis}
%---------------------------------------------------------------------------------------------------
%---------------------------------------------------------------------------------------------------
%---------------------------------------------------------------------------------------------------
%---------------------------------------------------------------------------------------------------
%----------------------------------------------------------------------------------------

Trend analysis exploits statistical methods to spot an underlying pattern in a
series of data (usually a time series), which can be distinguished from
randomness. 
It is widely used in disciplines such as economics, finance and environmental sciences.

\begin{figure} 
\centerline{\includegraphics[angle=0,width=8cm]{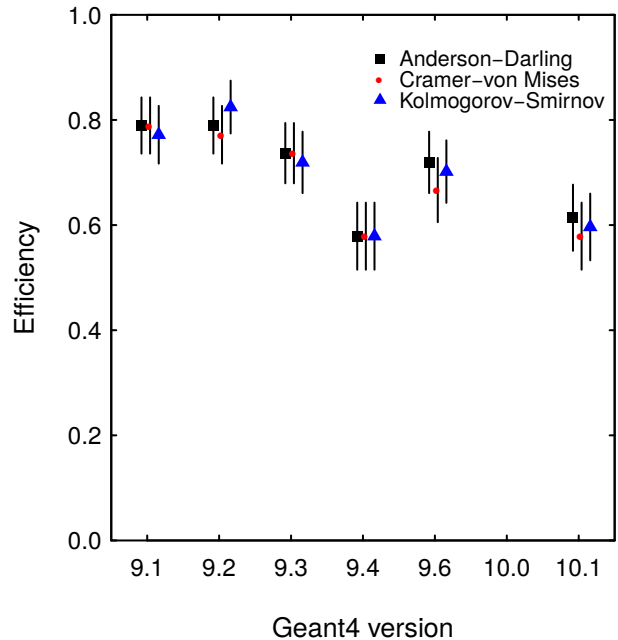}}
\caption{Evolution of compatibility with experiment (``efficiency'') of a simulation configuration versus 
Geant4 version. }
\label{fig_trend}
\end{figure}

The need of performing a trend analysis may also arise in physics software scenarios.
An example is illustrated in Fig. \ref{fig_trend}, which represents the
evolution of compatibility with experiment of a simulated observable (the
fraction of backscattered electrons, in this case), produced with the same user
application code \cite{tns_ebscatter1, tns_ebscatter2, tns_ebscatter3}, but using different Geant4 versions.
If the reference experimental data are unchanged, one expects that compatibility
with experiment would remain unchanged (within statistical fluctuations) over
different Geant4 versions, or at most would improve with time, if 
Geant4 itself is improved in later versions.
Otherwise, evolution of the observable towards worse compatibility with
experiment could be attributed to deterioration of the Geant4 kernel.
Fig. \ref{fig_trend} qualitatively hints to some apparent 
downward trend.
In such a scenario one requires the ability to discern whether the apparent
degradation of compatibility with experiment is statistically significant, so
that it would justify appropriate actions both by users (for instance, sticking
to an older version of Geant4 for production) and by Geant4 developers
(investigating the introduction of defects Geant4 code).

Various statistical analysis methods are available for trend analysis. 
We investigated the use of the Mann-Kendall test \cite{mann, kendall}, which tests whether to reject the
null hypothesis (H$_0$: no monotonic trend) in favour of the alternative
hypothesis (H$_1$: monotonic trend is present, e.g. downward trend in the case of Fig. \ref{fig_trend}).

In the scenario of Fig. \ref{fig_trend} the p-value resulting from the
Mann-Kendall test is 0.007, which corresponds to rejecting with 0.01
significance the null hypothesis that the observed pattern is consistent with
randomness, in favour of the alternative hypothesis of some degradation in compatibility with experiment.

It is worthwhile to note that trend analysis over an extended range can
identify more subtle effects than the mere comparison of just two scenarios
(e.g. in the aforementioned examples, of the outcome of simulation with two 
Monte Carlo simulation versions only), where genuine differences could be hidden by statistical
fluctuations.
This analysis technique could be a powerful instrument to complement 
validation and regression testing of Monte Carlo simulation systems, 
as well as in other experimental application scenarios.

\begin{figure} 
\centerline{\includegraphics[angle=0,width=8cm]{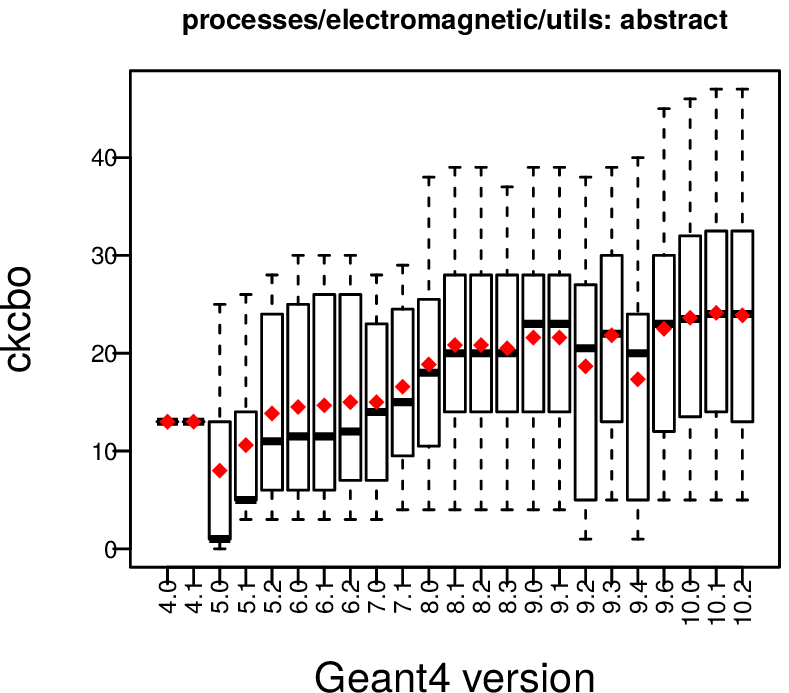}}
\caption{Evolution of the Coupling Between Objects software metric of abstract classes in the \textit{utils}  package of Geant4 electromagnetic physics versus 
Geant4 version. }
\label{ckcbo_abstract}
\end{figure}

\begin{figure} 
\centerline{\includegraphics[angle=0,width=8cm]{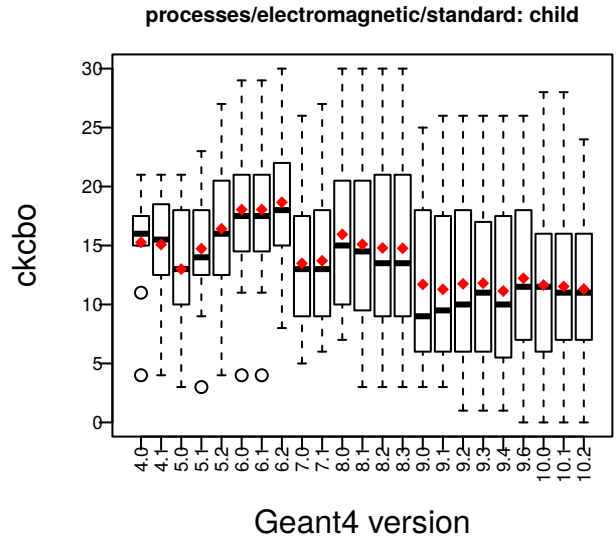}}
\caption{Evolution of the Coupling Between Objects software metric of leaf classes in the \textit{standard}  package of Geant4 electromagnetic physics versus 
Geant4 version. }
\label{ckcbo_child}
\end{figure}

We performed trend analysis also over a set of software quality metrics discussed in section 
\ref{sec_econom}. 
In this context we studied the evolution of metrics calculated over several Geant4 versions.
For instance, evolution towards greater coupling between objects is observed in Fig. \ref{ckcbo_abstract},
while a decreasing trend is observed in Fig. \ref{ckcbo_child}, which concern abstract base classes in 
the \textit{utils} and leaf (derived or non-derived) classes in the \textit{standard} packages of Geant4 electromagnetic physics, respectively.
In both cases the Mann-Kendall test rejects the hypothesis of randomness with 0.01 significance, in favour
of an alternative hypothesis of upward and downward trend, respectively.
Since various classes in the \textit{standard} package derive from abstract base classes in 
the \textit{utils} package, caution should be exercised in appraising the apparent decrease in 
coupling between objects observed in the  \textit{standard} package.

%---------------------------------------------------------------------------------------------------
%---------------------------------------------------------------------------------------------------
%---------------------------------------------------------------------------------------------------
%---------------------------------------------------------------------------------------------------
%---------------------------------------------------------------------------------------------------
\section{Conclusions}\label{sec: con}
%---------------------------------------------------------------------------------------------------
%---------------------------------------------------------------------------------------------------
%---------------------------------------------------------------------------------------------------
%---------------------------------------------------------------------------------------------------
%---------------------------------------------------------------------------------------------------

Data analysis methods pertinent to disciplines other than physics are powerful 
instruments in physics software applications.
Only a brief overview of our ongoing work in this innovative domain is given 
in this summary; more extensive results will be discussed in the conference presentation and in the full paper.

\label{------------------------------------------------END-------------------------------}

%\bibliography{nss}

\end{document}